\documentclass[aps,prl,amsmath,amssymb,showpacs,preprint,superscriptaddress]{revtex4-1}
\usepackage{graphicx}
\usepackage[latin1]{inputenc}
\usepackage[english]{babel}
\usepackage{color}
\usepackage{soul}

\newcommand{\com}[1]{}

\begin{document}




\title{Universal pinning energy barrier for driven domain walls in thin ferromagnetic films}


\author{V. Jeudy}
\email[]{vincent.jeudy@u-psud.fr}
\affiliation{Laboratoire de Physique des Solides, CNRS, Univ. Paris-Sud, Universit\'e Paris-Saclay, 91405 Orsay Cedex, France.}
\author{A. Mougin}
\affiliation{Laboratoire de Physique des Solides, CNRS, Univ. Paris-Sud, Universit\'e Paris-Saclay, 91405 Orsay Cedex, France.}
\author{S. Bustingorry}
\affiliation{CONICET, Centro At\'omico Bariloche, 8400 San Carlos de Bariloche, R\'{\i}o Negro, Argentina.}
\author{W. Savero Torres}
\author{J. Gorchon}
\affiliation{Laboratoire de Physique des Solides, CNRS, Univ. Paris-Sud, Universit\'e Paris-Saclay, 91405 Orsay Cedex, France.}
\author{A. B. Kolton}
\affiliation{CONICET, Centro At\'omico Bariloche, 8400 San Carlos de Bariloche, R\'{\i}o Negro, Argentina.}
\author{A. Lema\^itre}
\affiliation{Laboratoire de Photonique et de Nanostructures, CNRS, Universit\'e Paris-Saclay, 91460 Marcoussis, France.}
\author{J-.P. Jamet}
\homepage{deceased}
\affiliation{Laboratoire de Physique des Solides, CNRS, Univ. Paris-Sud, Universit\'e Paris-Saclay, 91405 Orsay Cedex, France.}

\email[]{vincent.jeudy@u-psud.fr}

\date{\today}

\begin{abstract}
We report a comparative study of magnetic field driven domain wall motion in thin films made of different magnetic materials for a wide range of field and temperature. The full thermally activated creep motion, observed below the depinning threshold, is shown to be described by a unique universal energy barrier function. Our findings should be relevant for other systems whose dynamics can be modeled by elastic interfaces moving on disordered energy landscapes.
\end{abstract}



\maketitle


\subsection{}
\subsubsection{}

\begin{figure}
\includegraphics[width=0.45\textwidth]{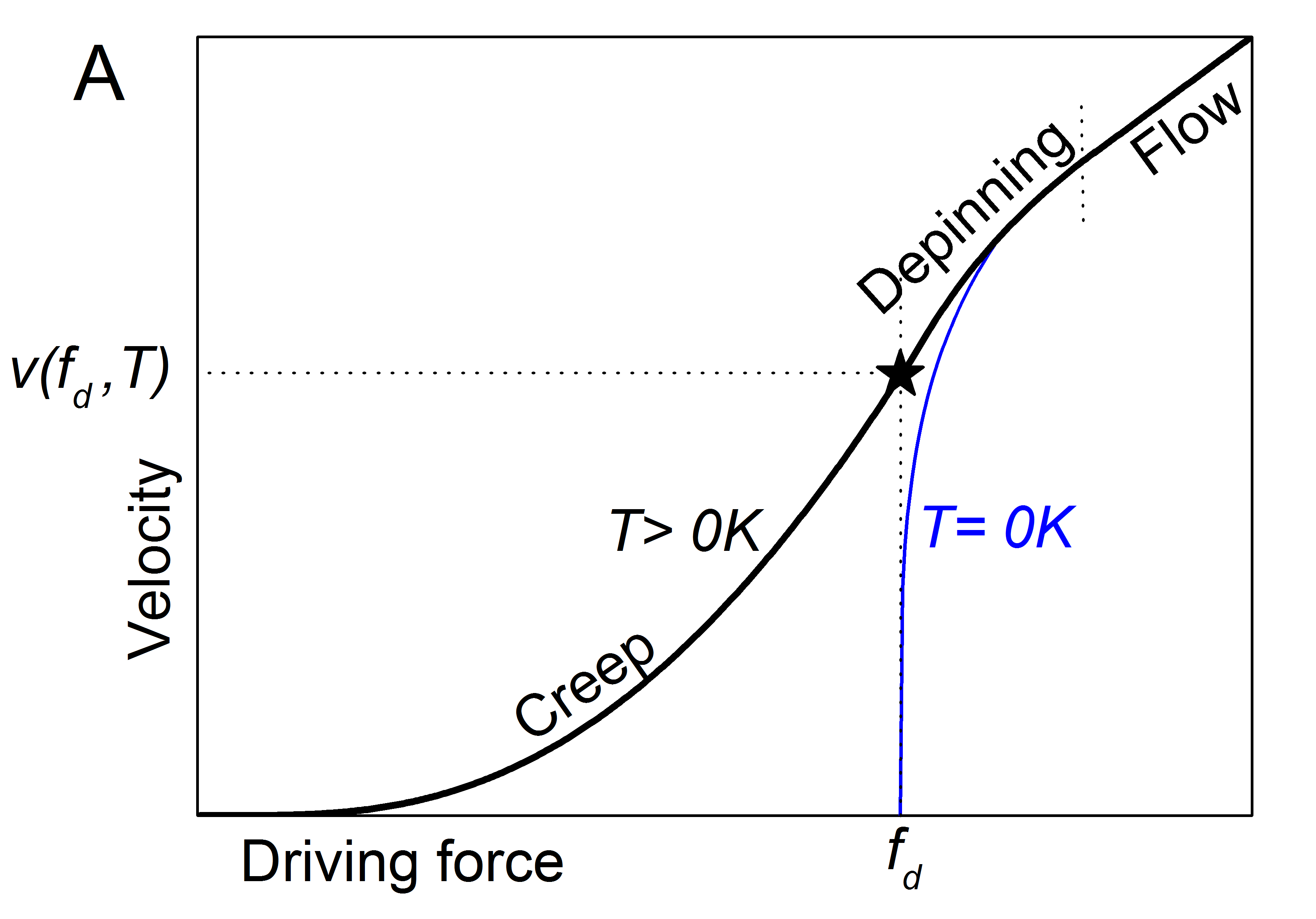}
\includegraphics[width=0.40\textwidth]{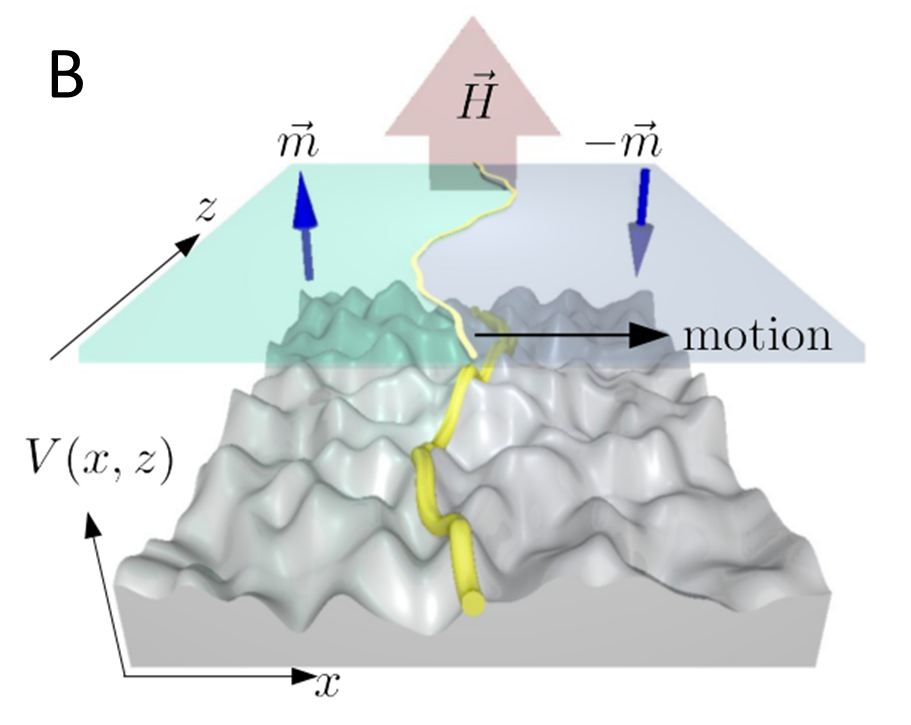}
\caption{\textbf{Motion of pinned elastic interfaces.} 
\textbf{A} Velocity-force characteristics for zero (blue curve) and finite (black curve) temperature showing the different dynamical regimes, the depinning threshold $f_d$ and the upper boundary (black star) of the thermally activated creep regime. \textbf{B} (Top view) The out-of-plane magnetic field $H$ favors the growth of up magnetization regions thus driving the domain wall (represented by the yellow spaghetti) in the right direction. (Bottom view) Theoretically, the creep domain wall dynamics in a thin film can be modeled by the displacement of a one-dimensional elastic line coupled to an effective two-dimensional random pinning energy landscape $V(x,z)$. 
}
\label{fig:Fig1}
\end{figure}

Domain walls are at the basis of future applications of high-density memories in ferromagnetic materials \cite{Parkin11042008}.
In this type of memories data bits are built up of magnetic domains with opposite magnetization varying in size and/or in position, and hence working memories implies the displacement of domain walls.
Noteworthy, even  weak random pinning due to local defects or inhomogeneities in the host materials is known to have a strong effect on domain walls~\cite{Ferre2013651}. Pinning tends to stabilize domain wall positions \cite{Krusin-Elbaum2001}, introduces stochasticity~\cite{Kim_PRL2003_avalanches}, induces domain wall roughness and dramatically modifies the driven dynamics at small field and current~\cite{lemerle_domainwall_creep,Moon_PRL_2013,DuttaGupta_Nat_Phys_2015}.
A fundamental understanding on how weak disorder affects the dynamics of domain walls is thus critical for applications. This question is also particularly relevant on a wider context since pinning dependent motion of elastic interfaces is observed in a large variety of other systems such as ferroelectric materials \cite{tybell2002}, contact lines in wetting \cite{ledoussal_contact_line}, crack propagation \cite{bonamy_crackilng_fracture}, and earthquake models~\cite{jagla_kolton_earthquakes}. In all those systems, the competition between the interface elasticity and pinning leads to rich and complex thermally activated motion over effective energy barriers (see Fig. \ref{fig:Fig1}) described by universal law~\cite{Fisher_review1998,Yamanouchi21092007,Ferre2013651}.
Although remarkable efforts have been made in the last decades, a quantitative description of the thermally activate regime of slow motion, so-called creep regime, has remained elusive.

An essential starting point to seize the universal character of the pinning dependent motion is the zero temperature behavior (see Fig. \ref{fig:Fig1}A). For an elastic line driven by a force $f$, a depinning threshold $f_d$ separates a zero velocity regime for $f<f_d$ from a finite velocity regime for $f>f_d$.
A finite temperature value $T$ results in a thermally activated subthreshold creep motion with the velocity following an Arrhenius law $v\sim \exp(-\Delta E/k_B T)$.
For a near zero driving force ($f \to 0$), the motion of an elastic line requires to overcome a universal divergent energy barrier presenting a power law variation $\Delta E\sim f^{-\mu}$. The universal creep exponent $\mu$ presents a good agreement~\cite{lemerle_domainwall_creep,metaxas_depinning_thermal_rounding,Kim_IEEE_2009_creep_temp,Kim2009,Burrows_CoFeB_APL2013,Gorchon2014,Beach_JMMM_2015_general,DuttaGupta_Nat_Phys_2015} with the predicted value ($\mu=1/4$), for an elastic line moving in an uncorrelated disordered potential~\cite{chauve2000}. 
A more stringent test of universality demonstrating a compatibility between universal scaling exponents and dimensionality~\cite{chauve2000} was performed in ferromagnetic Pt/Co/Pt ultrathin films~\cite{lemerle_domainwall_creep,metaxas_depinning_thermal_rounding} but was not yet reproduced for other ferromagnetic materials or other physical systems.
Moreover, apart from the near zero drive creep ($f \to 0$), the situation is much less clear and the nature of the motion is not well established. 
Only numerically, minimal models~\cite{kolton_dep_zeroT_long} can be used to test the universal character of domain wall motion. The experimental investigations often go beyond the depinning transition~\cite{Gorchon2014} ($f \gtrsim f_d$) and even reach the non-universal flow regimes independent of pinning and limited by dissipation \cite{Parkin_NatPhys_2007_V_H,metaxas_depinning_thermal_rounding,bustingorry_thermal_rounding_prb,dourlat2008} ($f \gg f_d$). However, thermally activated regimes with velocity-laws different from the predicted creep law ($\Delta E\sim f^{-1/4}$)~\cite{Yamanouchi21092007,Kim_IEEE_2009_creep_temp,Gorchon2014,Beach_JMMM_2015_general} can be encountered. 
Hitherto, the issue of the universality of force-driven elastic line creep motion for different magnetic materials and other systems remains open.

In this work, we address the question of universality from a study of magnetic field driven creep motion in magnetic films with perpendicular anisotropy. In magnets, the elastic interfaces are magnetic domain walls and the force $f$ is proportional to the applied field $H$ (see Fig. \ref{fig:Fig1}B).
The field-induced domain wall dynamics is studied in a single-crystalline (Ga$_{0.95}$,Mn$_{0.05}$)(As$_{0.9}$,P$_{0.1}$) semi-conductor~\cite{Dietl_Ohno_Review_2014} and in an amorphous [Tb/Fe]$_5$ ferrimagnetic~\cite{PommierJMMM1994} multilayer over a large range of temperatures. 
The observations are compared with results published in the literature for ultrathin metallic films (Pt/Co/Pt \cite{metaxas_depinning_thermal_rounding,cormierJPD2011,Gorchon2014}, Au/Co/Au \cite{kirilyuk_AuCoAu}, and CoFeB(\cite{Burrows_CoFeB_APL2013}) which have polycrystalline or amorphous structures. 
All these materials present specific structural and magnetic properties which results in different domain wall elastic energies and random pinning properties. Regardless of those differences, our comparative study shows that the creep domain wall dynamics presents universal behavior.

\begin{figure}
\includegraphics[width=0.45\textwidth]{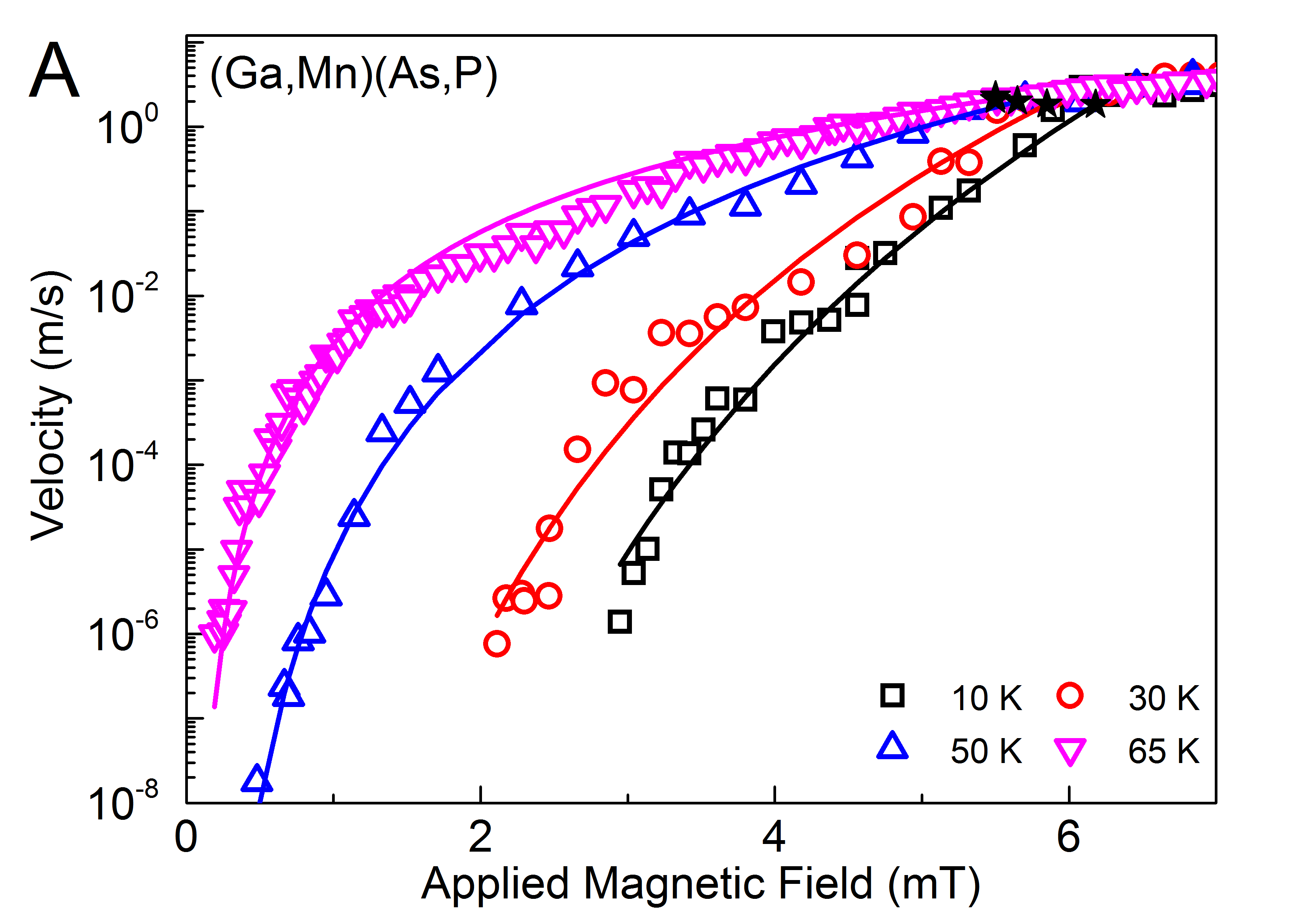}
\includegraphics[width=0.45\textwidth]{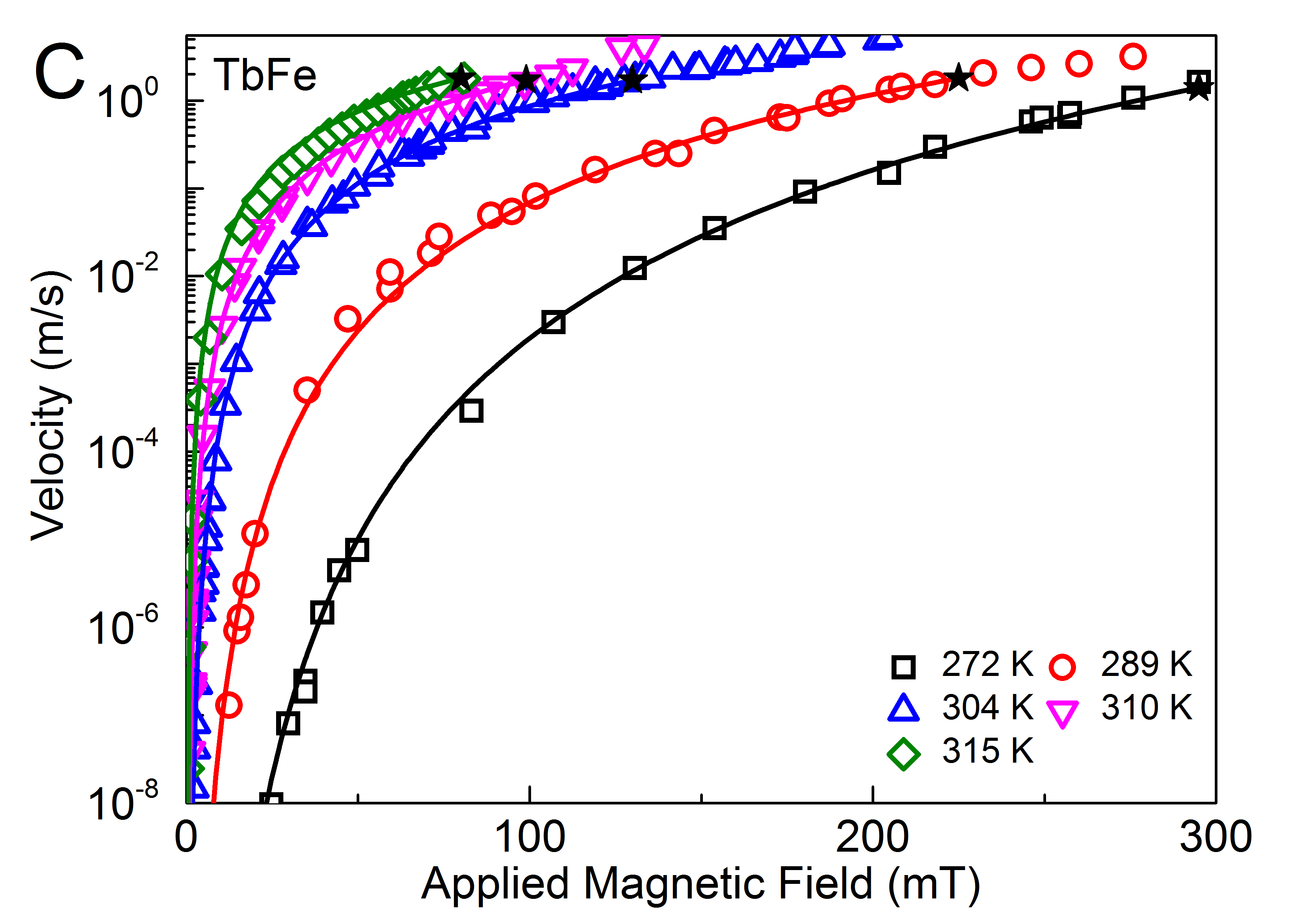}
\includegraphics[width=0.45\textwidth]{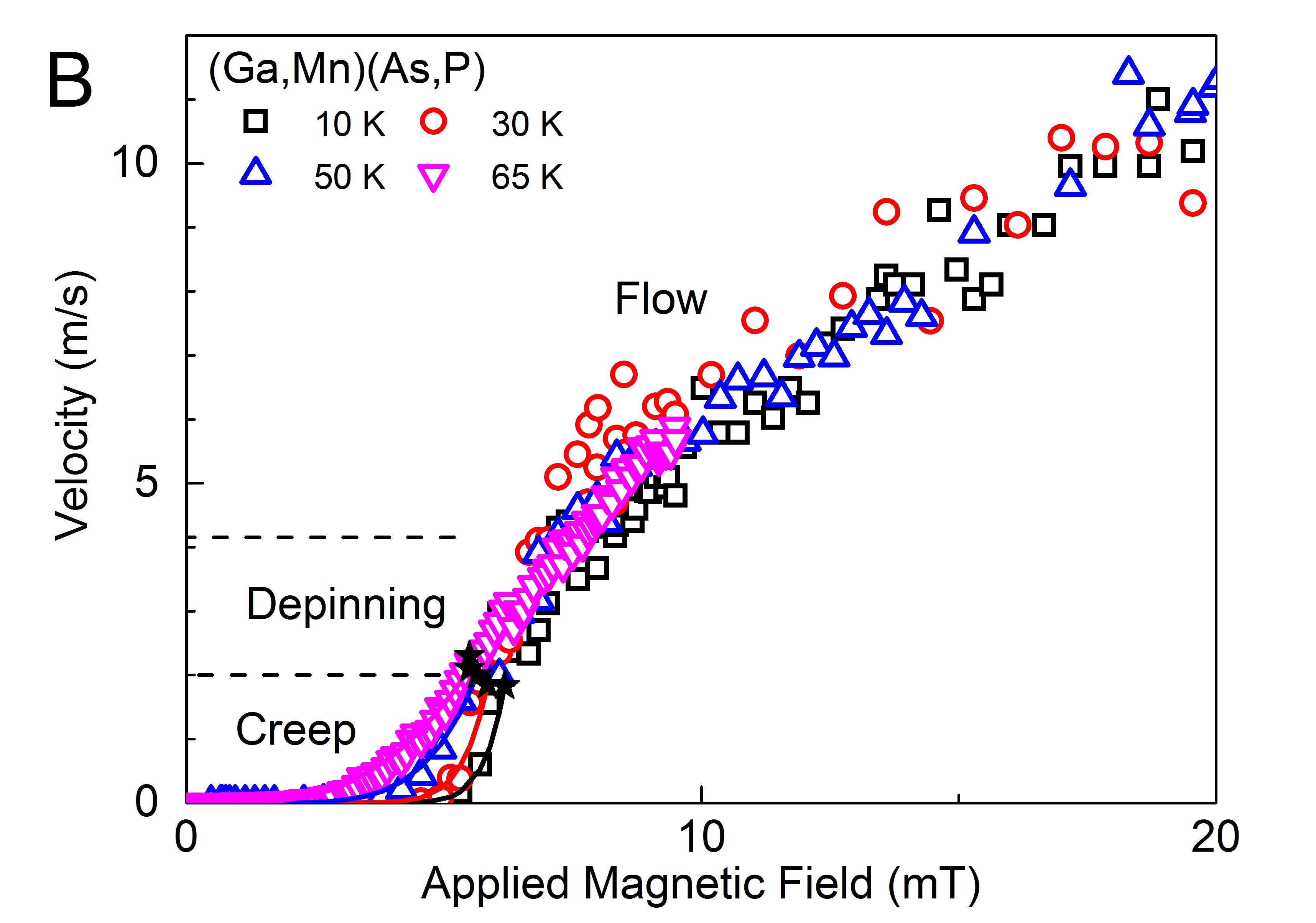}
\includegraphics[width=0.45\textwidth]{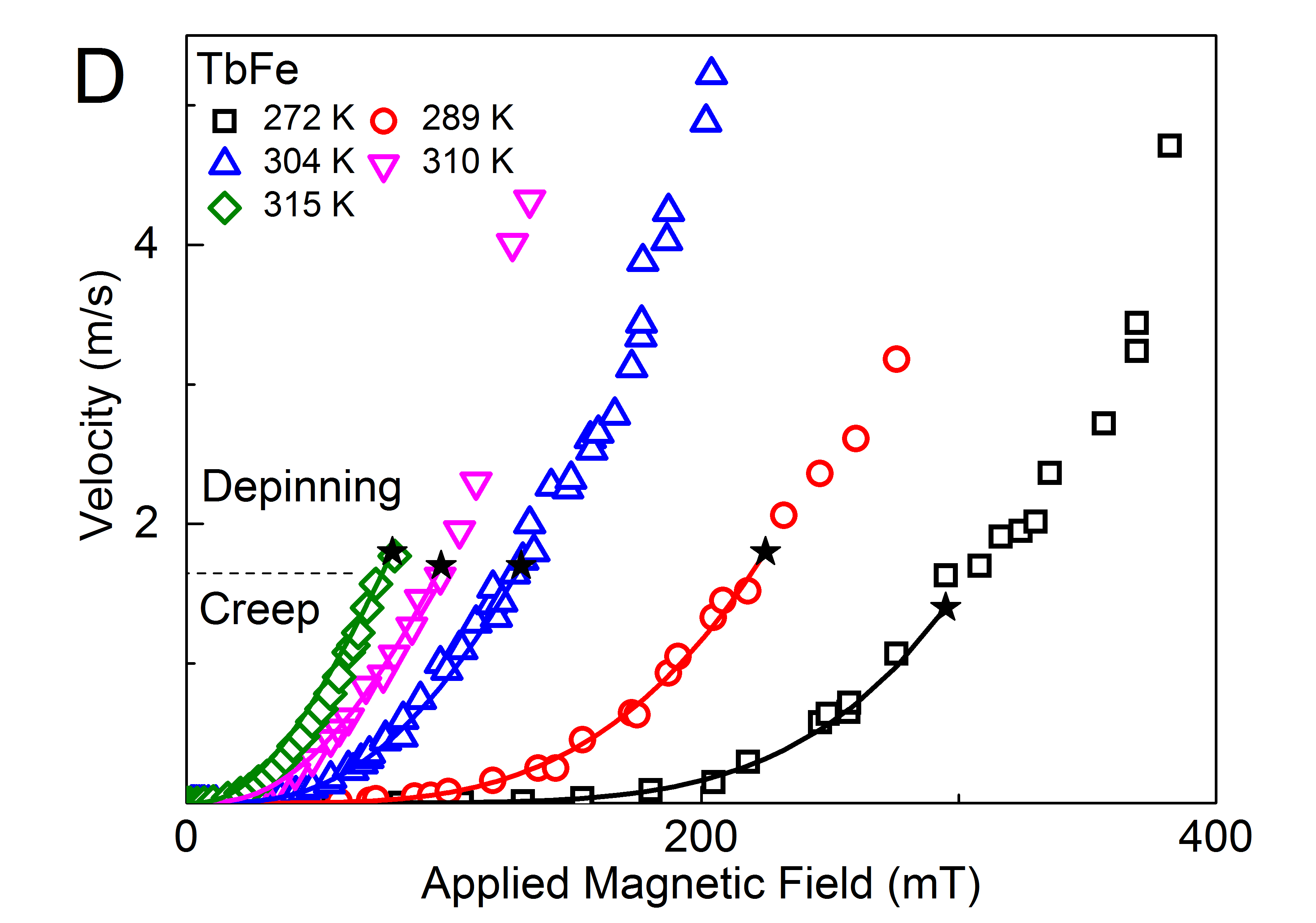}

\caption{\label{fig:Fig2} 
\textbf{Domain wall dynamics driven by a magnetic field.}
Velocity-field characteristics for a (Ga,Mn)(As,P) (curves \textbf{A} and \textbf{B}) and a TbFe (curves \textbf{C} and \textbf{D}) films measured over wide temperature ($10-65\mathrm{K}$ for (Ga,Mn)(As,P) and $272-315\mathrm{K}$ for TbFe) and magnetic field ranges.
The creep regime is highlighted by a semi-log plot (curves \textbf{A} and  and \textbf{C}). For both films, the velocity presents strong field and temperature variations, as expected for a thermally activated process. The linear flow regime is clearly observed at large field values ($H>10\mathrm{mT}$) for the (Ga,Mn,)(As,P) film (curves \textbf{B}) while it is beyond experimental reach for TbFe (curves \textbf{D}). 
The solid lines are fits of the creep law (using Eqs.~\ref{eq:v-creep} and \ref{eq:E-creep}) and the black stars correspond to the cross-over between creep and depinning regimes (coordinates: $H_d, v(H_d)$).
}
\end{figure}

The (Ga$_{0.95}$,Mn$_{0.05}$)(As$_{0.9}$,P$_{0.1}$) film is a 12~nm thick semi-conducting ferromagnetic single crystal. It was grown by low-temperature ($T=~250^\circ$C) molecular beam epitaxy on a GaAs~(001) substrate and was then annealed at $T= 200 ^\circ$C, for 4~h in air. Its Curie temperature is 74$\pm$1~K.
The multilayer film Si$_3$N$_4$(11nm)/[Tb(0.8nm)/Fe(1nm)]$ \times 5$ /Si$_3$N$_4$(11nm) have been deposited on floated glass substrate using a reactive diode rf sputtering system. Pure Tb, Fe, and Si targets were used and the pressure of 8 mTorr  was regulated in the chamber. The Si$_3$N$_4$ layer was obtained by pulsing a nitrogen partial pressure flux maintained at 2 mTorr near the samples during the silicon deposition. The Curie temperature is around $340\mathrm{K}$ and no compensation is observed in the $270-340\mathrm{K}$ temperature range. In order reach different temperatures, we used an open cycle He cryostat for the (Ga,Mn)(As,P) film and a home-made variable temperature system for the TbFe film. The motion domain walls was observed in a MOKE microscope. Their displacement was produced by magnetic field pulses of adjustable amplitude and duration ($1 \mu s-1s$) applied perpendicularly to the films. The domain wall velocity is calculated by the ratio between the displacement and the pulse duration.
 Velocity-field characteristics observed for the (Ga,Mn)(As,P) and the TbFe films, reported in Fig. \ref{fig:Fig2}, are good illustrations of typical results reported in the literature for the creep, depinning and flow regimes~\cite{lemerle_domainwall_creep,metaxas_depinning_thermal_rounding,Kim_IEEE_2009_creep_temp,dourlat2008,Yamanouchi21092007,Beach_JMMM_2015_general}.

\com{(Creep regime)} 
At low drives ($H \leq H_d$, the depinning threshold  $H_d$ is indicated by black stars in Fig. \ref{fig:Fig2}), domain walls follow the \textit{creep motion} (see Figs. \ref{fig:Fig2}A and \ref{fig:Fig2}C). The velocity presents a strong dependency on magnetic field, varying over several orders of magnitude for a relatively narrow applied magnetic field range. Increasing temperature is found to shift the curves towards low field values, thus reflecting the strong contribution of thermal activation.
The \textit{flow regime} is characterized by a linear variation of the velocity which is seen at sufficient large drive ($H \gg H_d$) for the (Ga,Mn)(As,P) film (see Fig. \ref{fig:Fig2}B). This non-universal regime is controlled by material dependent microscopic dynamical structure of domain walls~\cite{dourlat2008}. The flow regime is only encountered  in materials presenting a sufficiently low depinning threshold, like Pt/Co/Pt~\cite{metaxas_depinning_thermal_rounding,Beach_JMMM_2015_general}, (Ga,Mn)As~\cite{dourlat2008,Yamanouchi21092007}, FeNi~\cite{Beach_NatMat2005_V_H}, Pt/Co/AlO$_x$\cite{Miron_Nat_Mat_2011}. For other materials as TbFe (see Fig. \ref{fig:Fig2}D) and Au/Co/Au (\cite{kirilyuk_AuCoAu}), the flow regime was never observed experimentally.
\com{(Depinning regime)} 
The \textit{depinning transition} is manifested at intermediate drive ( $H \gtrsim H_d$). The cross-over between creep and depinning is found to occur for a $H_d$-value two orders of magnitude higher for TbFe than for (Ga,Mn,)(As,P) (see Figs. \ref{fig:Fig2}B and \ref{fig:Fig2}D), thus reflecting strongly different material dependent pinning properties.

\com{\subsection{(Self-consistent description of domain wall dynamics, definition of vd,Hd, Td, and links to the graphs)}}

To go beyond this qualitative presentation, we propose to describe empirically the whole creep regime $[0<H<H_d(T)]$ by a velocity given by:
\begin{equation}
 v(H,T) = v(H_d,T) \exp\left(-\frac{\Delta E}{k_B T} \right)
 \label{eq:v-creep}
\end{equation}
with 
\begin{equation}
 \Delta E= k_B T_d \left[ \left( \frac{H}{H_d} \right)^{-\mu} - 1 \right],
 \label{eq:E-creep}
\end{equation}
where $k_B T_d$ is the characteristic pinning energy scale, and $k_B$ the Boltzmann constant. Note that a similar empirical law was proposed in various theoretical works~\cite{Muller_PRB_01,roters_creep}. It is easy to see that Eq. \ref{eq:E-creep} yields the creep law 
$\Delta E \sim k_B T_d (H/H_d)^{-\mu}$ for $H \to 0$ and a linear vanishing of the energy barrier ($\Delta E \to 0$) for $H\to H_d$~\cite{Muller_PRB_01}. 
%
Each velocity-field characteristic was fitted by Eqs. \ref{eq:v-creep} and \ref{eq:E-creep} (see~\cite{supplementary_material}) in order to determine the parameters $H_d$, $v(H_d,T)$, and $T_d$. 
As reported in the supplementary material~\cite{supplementary_material}, the obtained values are material and temperature dependent reflecting the microscopical origin of pinning ~\cite{lemerle_domainwall_creep,Gorchon2014}. However a discussion on this non-trivial subject goes beyond the scope of this letter.
As it can be observed in Fig.\ref{fig:Fig2} and in the Figs. of the supplementary material~\cite{supplementary_material}, the fitting of the velocity-field characteristics works particularly well for the whole creep motion for all temperatures and materials.

\begin{figure}
\includegraphics[width=0.99\textwidth]{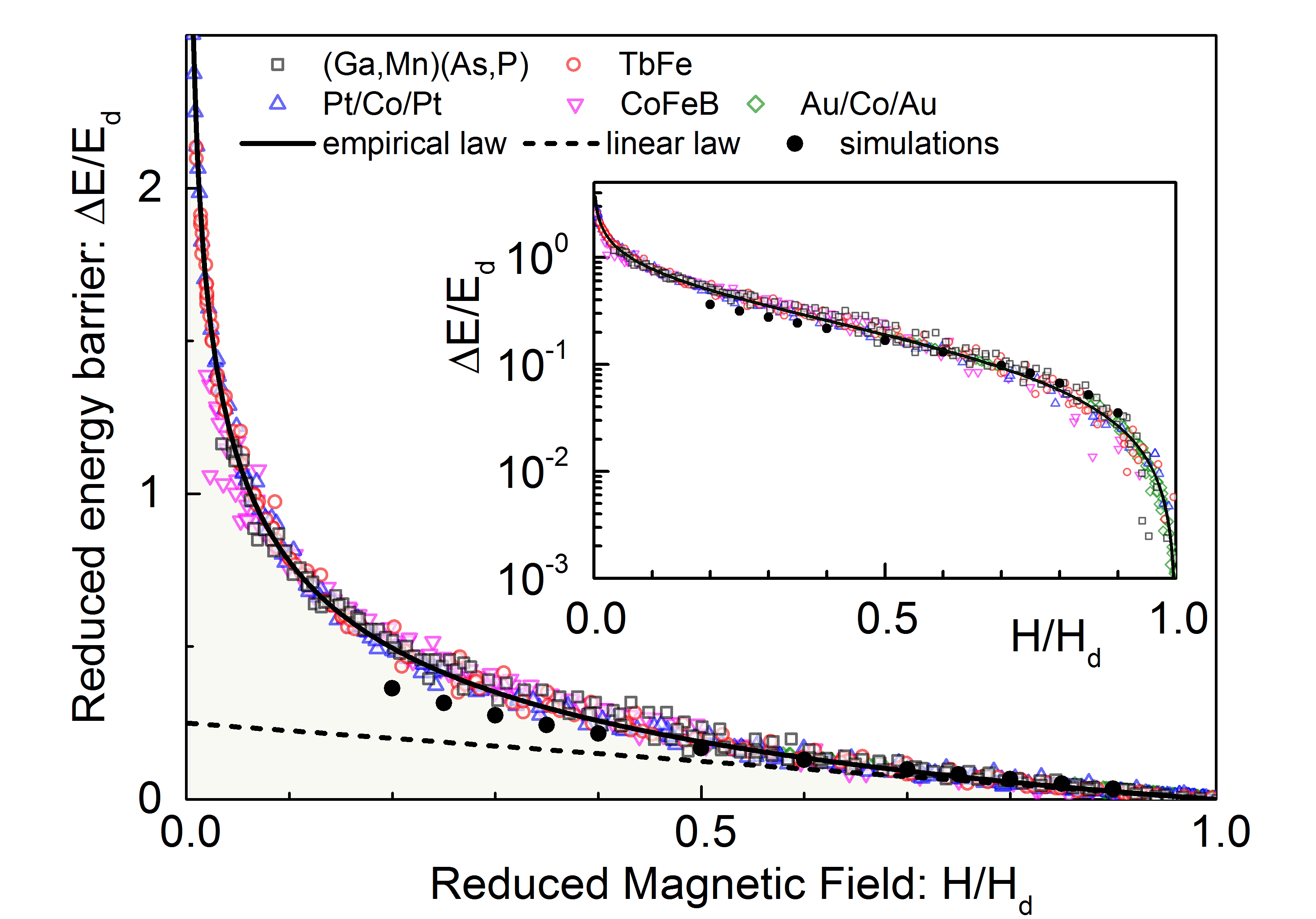}
\caption{\label{fig:Fig3}
\textbf{Universal energy barrier of the creep regime. }
The variation of the reduced energy barrier height $\Delta E/E_d$, with $E_d=k_B T_d$, is reported as a function of the reduced force $H/H_d$, for five different magnetic materials and for temperatures ranging from 10 to 315 K (25 curves are superimposed). The solid line is a plot of Eq.~\ref{eq:E-creep}. The black circles correspond the predictions of ref.~\cite{kolton_dep_zeroT_long} whose energy scale was adjusted to experimental data (see~\cite{supplementary_material}). The dashed line corresponds to the linear variation of the energy barrier close to the depinning field $(H = H_d)$. Inset: Universal barrier presented in semi-log scale showing a good quantitative agreement with Eq.~\ref{eq:E-creep} over more than three orders of magnitude. 
}
\end{figure}

\com{\subsection{(Universal energy barrier)}}
Let us now address the question of the universality of domain wall dynamics. 
\com{(\textit{methods: how the curve is obtained})} Once the material dependent parameters are obtained, Eq. \ref{eq:v-creep} can be used to deduce the pinning barrier height $\Delta E$ from the velocity curves $v(H,T)$, i.e., $\Delta E(H) = k_B T \ln[(v(H_d,T)/v(H,T)]$. In Fig. \ref{fig:Fig3}, the reduced energy barrier height $(\Delta E/k_B T_d(T))$ is plotted as a function of the reduced force $(H/H_d(T))$ for the (Ga,Mn)(As,P) and TbFe thin films, and three other ferromagnetic materials.   
\com{(\textit{Description of the curve and technical aspects}) }
As it can be observed, all the reduced energy barrier curves $\Delta E/k_B T_d(T)$ collapse well onto a single master curve over the full field-range of existence of the creep regime $0<H/H_d<1$. This is the central result of this letter. It demonstrates that for magnetic materials the \textit{creep} can be described by a unique energy barrier function, extending from the divergent barriers dominated dynamics close to $H=0$ to the regime where the barriers vanish linearly ($ \Delta E \sim (1/4) k_B T_d \left[H/H_d - 1 \right]$) approaching $H_d$. 
%
This result is very robust since the master curve is obtained for a variation of the reduced energy barrier heights (see inset of Fig. \ref{fig:Fig3}) over more than three orders of magnitude $(\Delta E/k_B T_d = 10^{-3} - 5)$ and for an explored range of temperature of almost two orders of magnitude $(T = 10-315\mathrm{K})$. 
\com{(\textit{Impact of this result in terms of understanding pinning dependent motion})}
The quantitative agreement between Eq.~\ref{eq:E-creep} which was postulated 15 years ago as a possible interpolation formula and the experimental data, demonstrates that the creep regime can be very well described by a unique (reduced) barrier function of the (reduced) temperature and field.
Furthermore, it compares fairly well with the results obtained by using numerical simulations~\cite{kolton_dep_zeroT_long} of a minimal model for a one-dimensional elastic line in a two-dimensional disordered medium (see Fig. \ref{fig:Fig3}).
As this model does not take into account the properties of a specific system, the barrier function of the creep regime is expected to be relevant for a larger variety of systems other than ferromagnets. 

\com{\section{Conclusion}}
In conclusion, we provide evidence of the universal character of the whole thermally activated subthreshold creep motion in magnetic thin films. In this dynamical regime, the magnetic domain wall motion is shown to be controlled by a unique universal reduced energy barrier function. 
The compatibility of this universal law with the predictions of a minimal model strongly suggests our results to be relevant to understand the creep dynamics in other systems than magnetic thin films whose emergent properties are also controlled by the competition between quenched disorder and the elasticity of a driven fluctuating string. 

\begin{acknowledgments}
We wish to thank N. Vernier and J.-P. Adam for providing us their data on CoFeB and J. Ferr\'e, A. Thiaville, and T. Giamarchi for fruitful discussions. S. B., J. G., A.M. and V. J. acknowledge support by the French-Argentina project ECOS-Sud num.A12E03. This work was also partly supported by the french projects DIM CNano IdF (Region Ile-de-France), the RENATECH network, and the Labex NanoSaclay, Ref.: ANR-10-LABX-0035. S. B. and A. B. K. acknowledge partial support from Projects PIP11220090100051 and PIP11220120100250CO (CONICET).
\end{acknowledgments}

\bibliography{refs_AM2}

\end{document}